\documentclass[prl, twocolumn, superscriptaddress]{revtex4-1}
\usepackage{amsmath}
\usepackage[english]{babel}
\usepackage[pdftex]{graphicx}
\usepackage{color}
\usepackage{ulem}
\usepackage{comment}

\usepackage[sort&compress]{natbib}
\usepackage{hyperref}
\graphicspath{{images/}{./}}

\usepackage{amsmath}
\usepackage{amssymb}
\usepackage{mathtools}
\usepackage{braket}

\renewcommand{\vec}[1]{\ensuremath{\boldsymbol{#1}}}

\renewcommand{\deg}{^{\circ}}

\begin{document}
\title{Identifying Klein tunneling signatures in bearded SSH lattices from bent flat bands}
\author{Yonatan Betancur-Ocampo}
\email{ybetancur@fisica.unam.mx}
\affiliation{Instituto de F\'isica, Universidad Nacional Aut\'onoma de M\'exico, Ciudad de México, Mexico}

\author{Guillermo Monsivais}
\email{monsi@fisica.unam.mx}
\affiliation{Instituto de F\'isica, Universidad Nacional Aut\'onoma de M\'exico, Ciudad de México, Mexico}

\begin{abstract}
    Su-Schrieffer-Heeger (SSH) lattices have helped to understand key concepts of topological insulators. In this paper, we study the transmission properties of $pn$ junctions of bearded SSH lattices forming a bipartite structure. From a tight-binding approach, a Bloch Hamiltonian depicts the electron behavior as a massive pseudo-spin one particle in quasi-one-dimensional systems. It is possible to design the band structure of these lattices by tuning the on-site energy and hopping parameters. The characteristic flat band of pseudo-spin one systems can be bent by the staggered potential of the bottom atom in the bearded edge. We explore interband transmissions from the bent flat band to the valence one of a bipartite structure, where Klein tunneling can be obtained depending on the hopping parameter associated with the bearded edge. The identification of Klein tunneling signatures is feasible in artificial lattices such as photonic lattices, elastic resonators, topolectric circuits, or quantum dot arrangements. On the other hand, the study of Klein tunneling in polyatomic molecules may contribute to advance in one-dimensional transistors.
\end{abstract} 

\maketitle

\section{Introduction}
Klein tunneling is one of the most outstanding transport effects with experimental realizations in graphene and artificial systems \cite{Katsnelson2006,Allain2011,Beenakker2008,CastroNeto2009,Young2009,Jiang2020}. This effect consists of the perfect crossing of a particle within the electrostatic potential profile, where kinetic energies can be less than the barrier potential height \cite{Klein1929,Katsnelson2006,Beenakker2008,CastroNeto2009}. Since 1929, Oskar Klein predicted this tunneling in the context of high-energy physics without testing so far \cite{Klein1929}. In condensed matter, Klein tunneling emerges when electrons pass from one band to another with the conservation of an associated pseudo-spin lattice \cite{Katsnelson2006}. From the experimental realization of Klein tunneling in graphene \cite{Young2009}, multiple proposals of related phenomena have been predicted with some tests in artificial lattices \cite{Jiang2020,Zhang2022c,Zhu2023}. On this recent topic is remarkable the prediction of super-Klein tunneling of electrons in two-dimensional lattices with unit cells containing three atoms, among them: dice lattices, Lieb lattices, and honeycomb superlattices \cite{Bercioux2009,Shen2010,Dora2011,Lan2011,Urban2011,Xu2016,Bercioux2017,BetancurOcampo2017,ContrerasAstorga2020,Cunha2020,Chen2019,Wang2020,Hao2021,Liu2022,Kim2019,Wang2022,Liu2023,Duan2023,Kim2019a,BetancurOcampo2018,Jakubsky2023,Zeng2022,Korol2018,CrastodeLima2020,Nandy2019,Kim2020,Filusch2020}. In such systems, electrons behave as particles with pseudo-spin one, where a flat band lies between two dispersive bands \cite{Bercioux2017}. The study of materials with flat bands has increased recently due to its outstanding effects on electronic properties \cite{Wang2021,Zhang2022,Wang2021a,Wang2020a,Li2022,Zhang2022a,Zhang2022b,Wen2023,BetancurOcampo2017a,NavarroLabastida2023}. Super-Klein tunneling has a recent experimental realization in phononic crystals \cite{Zhu2023}. Also, anomalous Klein tunneling, which is a non-resonant perfect transmission angularly dependent, appears for borophene 8-pmmn and uniaxially strained graphene \cite{Xie2019,BetancurOcampo2018,Huang2023,Xu2023}. This phenomenon has experimental observation in photonic graphene \cite{Zhang2022c}. In Kekulé graphene superlattices, other type of Klein tunneling is predicted, where perfect transmission goes along with a valley-flip \cite{Garcia2022}.  Anti-Klein tunneling, which is perfect back-scattering under normal incidence, and its omnidirectional perfect reflection named anti-super-Klein tunneling, have been studied in 2D materials such as bilayer graphene and phosphorene \cite{Katsnelson2006,BetancurOcampo2019,BetancurOcampo2020,Majari2023,Septembre2023,Cunha2020,MolinaValdovinos2022}. Beyond the Klein tunneling, more phenomena with flat bands have been predicted, such as Andreev reflection \cite{Beenakker2008,Feng2020,Zeng2022,Septembre2023}, topological charge pump \cite{Wang2021}, super skew scattering \cite{Wang2021a}, Seebeck and Nernst effects \cite{Duan2023}, enhanced magneto-optical effect \cite{Chen2019}, Anderson localization \cite{Kim2019}, and electron-beam collimation \cite{Wang2022,Xu2016}.

Recently, there has been a remarkable increment of research related to one-dimensional lattices, particularly in artificial systems, where toy models based on the tight-binding approach are developed \cite{CaceresAravena2022,Coutant2021,Dietz2018,Torrent2013,MartinezArgueello2022,Stegmann2017,RamirezRamirez2020,Casteels2016,Majari2021,Belopolski2017,Torrent2013,Freeney2022,RamirezRamirez2020,Dietz2018,MartinezArgueello2022,Liu2022,Suesstrunk2015,Bellec2013,Drost2017}. Most of these investigations focus on the study of topological insulators \cite{Hasan2010,Asboth2016,Shen2017}. Photonic \cite{CaceresAravena2022}, phononic \cite{Coutant2021,Wang2020b,Li2018,Kim2020,Tang2022}, elastic wires \cite{Thatcher2022}, water waves \cite{Yang2016}, atomic arrangements \cite{Meier2016}, and quantum dots lattices \cite{Kiczynski2022} serve to emulate 1D chains of polyacetylene, veryfing topological states predicted by the Su-Schrieffer-Heeger (SSH) model \cite{Su1979,Heeger1988}. This model gives the most simple description of topological insulators, where outstanding effects have been studied, among them the formation of solitons \cite{Su1979} and the conductivity in polymers \cite{Heeger1988,Paasch1992}. Multiple extensions of the SSH model have been proposed \cite{Li2022a,Xie2019,Ahmadi2020,Kim2020,Lieu2018,Li2014,Manda2023}, which depicts unusual physical properties such as non-hermitian skin effect \cite{Liu2022}, Dirac states \cite{Li2022}, doublons \cite{Azcona2021}, and gap solitons \cite{Tang2022}. Likewise, the emergence of flat bands is not exclusive to higher dimensional lattices but also in one-dimensional chains, where localized states are observed \cite{Hao2023,Maimaiti2017,Huda2020,Catarina2022,Liu2013,Tilleke2020,Mukherjee2015}.

Nevertheless, Klein tunneling has been explored scarcely in 1D chains. It could be due to the difficulty of creating an abrupt $pn$ junction in polyacetylene and other molecules to test perfect tunneling. However, the rise of artificial lattices allows us to export unusual phenomena from condensed matter towards optics, acoustics, elastic waves, or another field \cite{Freeney2022}. Therefore, toy models based on a tight-binding approach may go beyond academic curiosity due to the universality of the Bloch theorem.

In this paper, we look for the parameters regime to obtain Klein tunneling in bearded SSH lattices. SSH chains are modified by adding extra bonds to create a bearded edge, as shown in Fig. \ref{Bearded_SSH_lattice}. The developed tight-binding model to nearest neighbors is applied to the three lattices shown in Fig. \ref{Bearded_SSH_lattice}. The main advantage for considering bearded edge is that excitations possess pseudo-spin one to the difference of the conventional SSH chains, where the electrons or quasi-particles have pseudo-spin 1/2. The most general situation to analyze the effects on the band structure is to take the three on-site energies and the hopping parameters of the unit cell differently. We establish the condition for creating a flat band in bearded SSH lattices. We analyze the bending effect in the flat band on the transmission in bipartite bearded SSH lattices to identify Klein tunneling signatures. Possible experimental setups to test the present results may be the implementation of elastic resonators, photonic chains, topolectric circuits, or quantum dot arrangements \cite{Freeney2022}.

\section{Tight binding approach of bearded SSH lattices}
\begin{figure}
    \centering
    \begin{tabular}{c}
    (a) \qquad \qquad \qquad \qquad \qquad \qquad \qquad \qquad \qquad \qquad \qquad \qquad \\ 
    \includegraphics[scale=0.32]{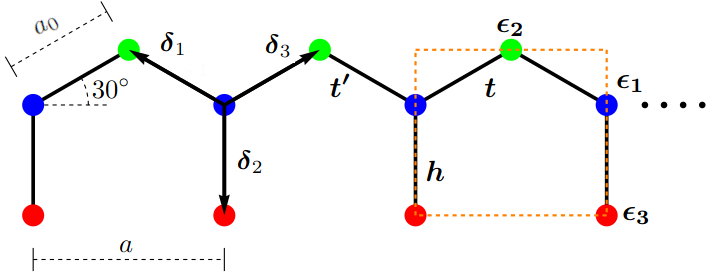}\\
    \includegraphics[scale=0.2]{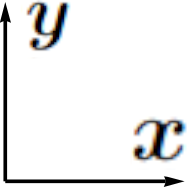}\\  
    (b) \qquad \qquad \qquad \qquad \qquad \qquad \qquad \qquad \qquad \qquad \qquad \qquad \\ 
    \includegraphics[scale=0.45]{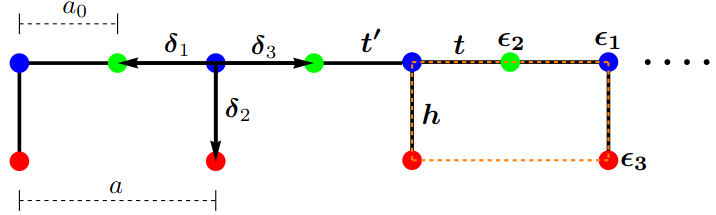}\\
     (c) \qquad \qquad \qquad \qquad \qquad \qquad \qquad \qquad \qquad \qquad \qquad \qquad \\ 
    \includegraphics[scale=0.45]{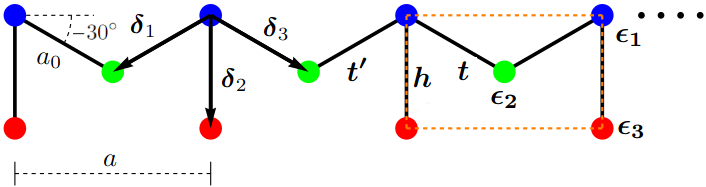}
    \end{tabular}
    \caption{Different geometries for the bearded Su-Schrieffer-Heeger (SSH) lattices. The atoms are indicated by colored circles and nearest bonds with black lines. The unit cell, which is drawn by a dashed orange rectangle, has a length of $a = 2a_0\cos\alpha$, and the relative atomic positions are given by $\vec{\delta}_1 = a_0(-\cos\alpha,\sin\alpha)$, $\vec{\delta}_2 = -a_0(0,1)$, and $\vec{\delta}_3 = a_0(\cos\alpha,\sin\alpha)$. Examples of bearded SSH lattices are the bearded edge of a zigzag graphene nanoribbon, where $\alpha = 30\deg$ in (a), the edge of a Lieb nanoribbon with $\alpha =0\deg$ in (b), while negative values of $\alpha$ can be considered, as shown in (c) with $\alpha = -30\deg$.}
   \label{Bearded_SSH_lattice}
\end{figure}

Bearded-SSH lattices are chains that possess a part identical to those described by the SSH model and a bearded edge whose unit cell has three atoms, as shown in Fig. \ref{Bearded_SSH_lattice}. Geometrically, these lattices can be different by changing the angle $\alpha$ in the zigzag bond. However, bearded SSH lattices are topologically identical because it is always possible to define the lattice constant as $a = 2a_0\cos\alpha$, where $a_0$ is the bond length and $\alpha$ is the bond angle for the zigzag edge. The tuning of $\alpha$ causes expansion of the band structure only with respect to the case $\alpha = 0$. We can consider three atomic positions within the unit cell with different on-site energies. In the tight-binding approach, we take into account the nearest-neighbor interactions only. The on-site energies in Fig. \ref{Bearded_SSH_lattice} are $\epsilon_1$, $\epsilon_2$, and $\epsilon_3$ for the atoms in blue, green, and red color, respectively. While hopping parameters $h$, $t$, and $t'$ describe the interaction to nearest neighbors between red-blue, blue-green, and green-blue atoms, respectively.

\begin{figure*}
    \centering
    \begin{tabular}{ccc}
    (a) \qquad \qquad \qquad \qquad \qquad \qquad \qquad \qquad & (b) \qquad \qquad \qquad \qquad \qquad \qquad \qquad \qquad & (c) \qquad \qquad \qquad \qquad \qquad \qquad \qquad \qquad\\
    \includegraphics[scale=0.265]{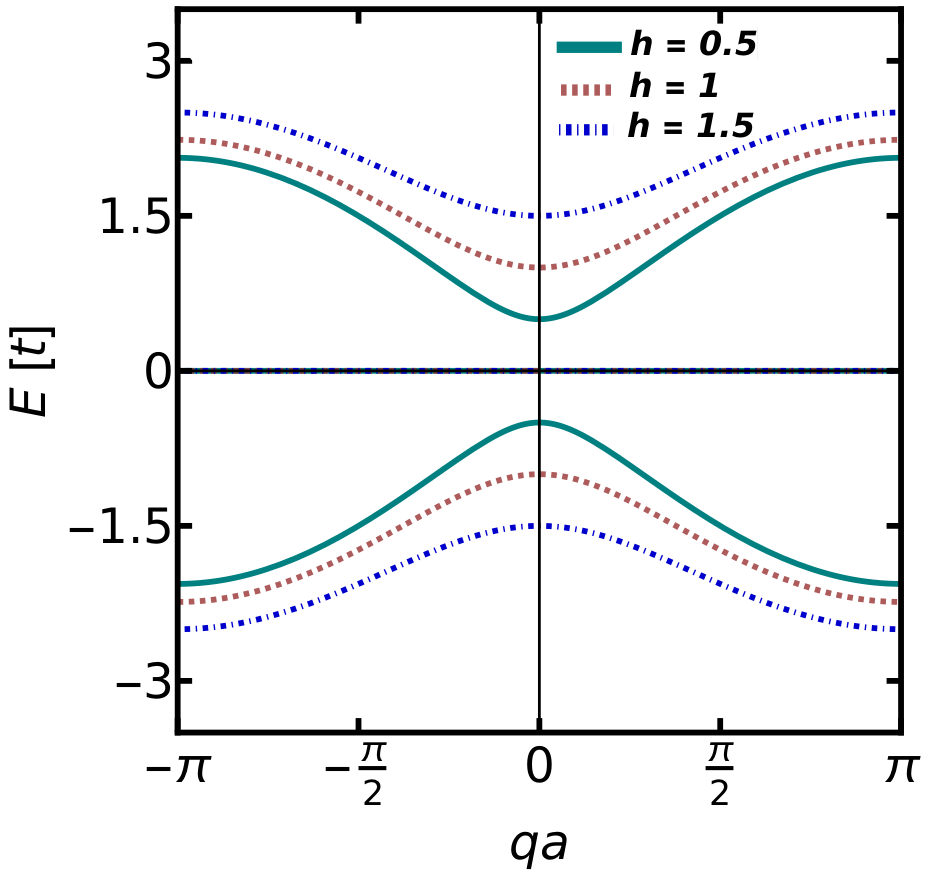}&
     \includegraphics[scale=0.265]{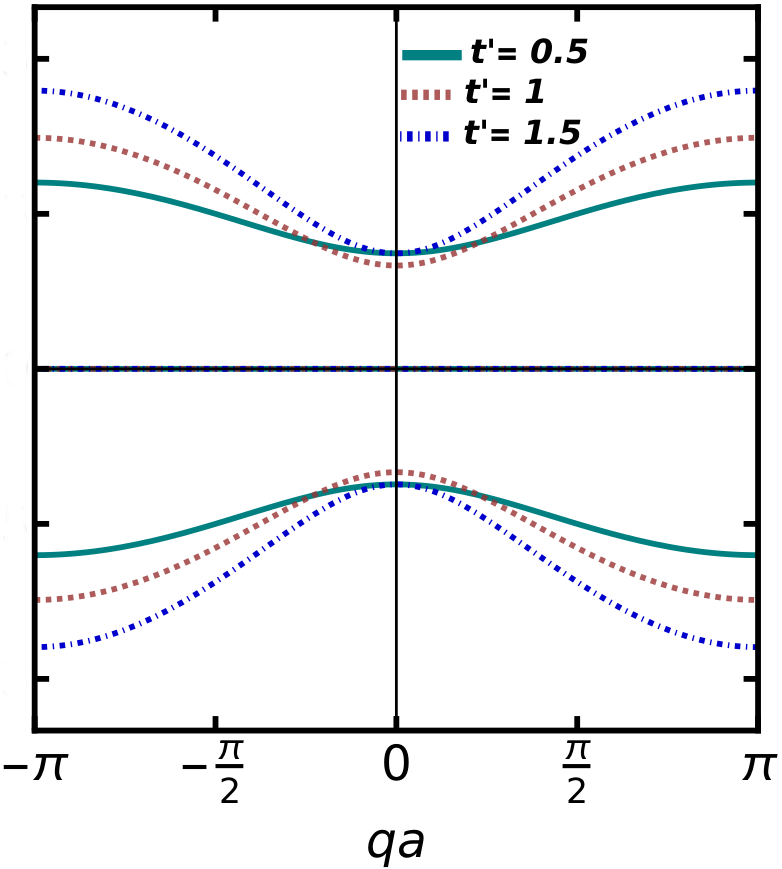}&
      \includegraphics[scale=0.265]{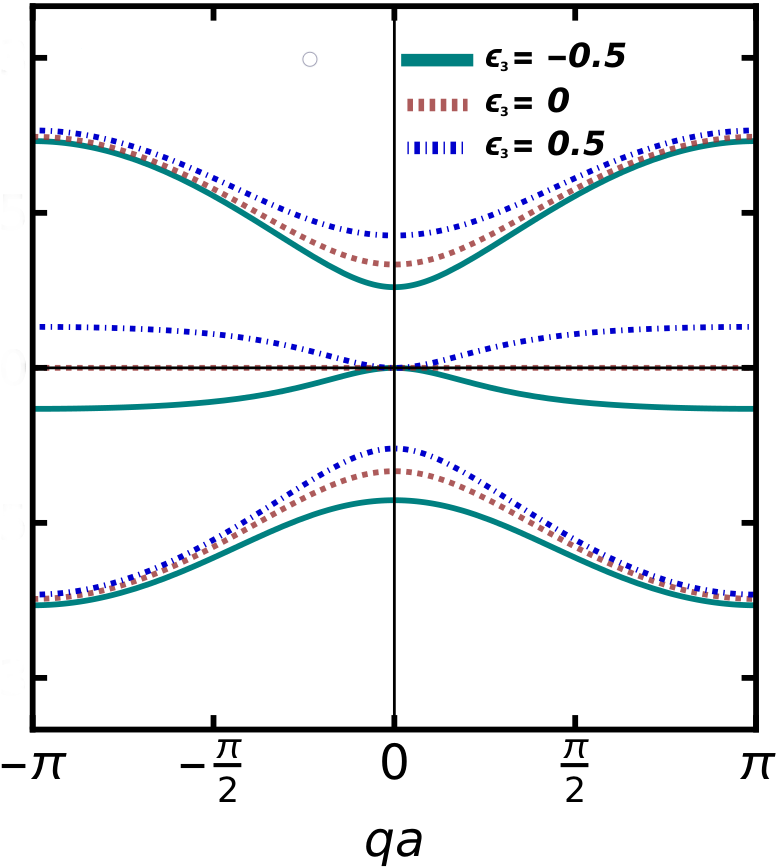}
    \end{tabular}
    
    \caption{(a) Band structure of the bearded SSH lattice. The colored curves correspond to different values of the hopping parameter $h$, where on-site energies are $\epsilon_i = 0$, with $i = 1, 2$, and 3, and the hopping parameters of the SSH model are $t = t' = 1$. The flat band remains unaffected by the change of $h$. (b) If we set the parameters $h = t = 1$ and $t' = 0.5, 1,$ and $1.5$, the curvature of the dispersive bands is tuned only with a constant flat band. (c) In the case $h = t = t' = 1$, $\epsilon_1 = \epsilon_2 = 0$, and by modifying the on-site energy $\epsilon_3 = -0.5, 0,$ and 0.5, the middle band lets to be flat when $\epsilon_2 \neq \epsilon_3$.}
   \label{Band_struct}
\end{figure*}

With these considerations, the Bloch Hamiltonian of the bearded-SSH lattice can be represented by a $3 \times 3$ matrix given by (see appendix)

\begin{equation}\label{HSSSH}
    H(k)= \left(\begin{array}{ccc}
        \epsilon_1 & g^*(k) & h \\
        g(k) & \epsilon_2 & 0 \\
        h & 0 & \epsilon_3
    \end{array}\right),
\end{equation}

\noindent where $k$ is the wave vector and $g(k) = t\exp(-ika/2)+t'\exp(ika/2)$. This Hamiltonian is similar to the Hamiltonian of the SSH model. Diagonalizing this Hamiltonian, we obtain three bands as a function of $k$. A conventional SSH model has two energy bands. The hopping parameter $h$ modulates the properties of the SSH chain with the introduction of an extra energy band, as shown in Fig. \ref{Band_struct}. While the on-site energy $\epsilon_3$ controls the bending of the middle band by offering a grade of freedom for the transmission.

One can relate the Hamiltonian in Eq. \eqref{HSSSH} with the spin-one matrices. This relation shows the behavior of the electrons like quasi-particles of pseudospin equal to one, as follows

\begin{equation}\label{HSSSH2}
    H(k) = 2\vec{S}\cdot\vec{g}(k) + M,
\end{equation}

\noindent where $\vec{S}=(S_x,S_y)$ are the spin one matrices given by

\begin{eqnarray}
    S_x = \frac{1}{2}\left(\begin{array}{ccc}
     0 & 1 & 0 \\
     1 & 0 & 0 \\
     0 & 0 & 0
    \end{array}\right) & \textrm{and} & S_y = \frac{1}{2}\left(\begin{array}{ccc}
     0 & -i & 0 \\
     i & 0 & 0 \\
     0 & 0 & 0
    \end{array}\right),
\end{eqnarray}

\noindent $\vec{g}(k)=(\textrm{Re}\{g(k)\},\textrm{Im}\{g(k)\})$, and $M$ is defined as a matrix responsible of the gap opening

\begin{equation}\label{M}
    M= \left(\begin{array}{ccc}
        \epsilon_1 & 0 & h \\
        0 & \epsilon_2 & 0 \\
        h & 0 & \epsilon_3
    \end{array}\right).
\end{equation}

\noindent Exact analytical expressions for the eigenenergies of the Hamiltonian in Eq. \eqref{HSSSH} are obtained by setting the on-site energy $\epsilon_2 = \epsilon_3$, and without loss of generality, doing $\epsilon_1 = -\epsilon_2 = \epsilon$. These eigenenergies are

\begin{equation}
    E_0 = -\epsilon \qquad \textrm{and} \qquad E_s = s\sqrt{\epsilon^2 + h^2+|g(k)|^2},
\end{equation}

\noindent where $s = \pm 1$ is the band index for the dispersive bands $E_s$ and $E_0$ is the energy for the flat band. The condition $\epsilon_2 = \epsilon_3$ gives always a flat band. Without this condition, the flat band becomes a heavy band, whose curvature is controlled by the hopping parameter $\epsilon_3$. 

It is possible to engineer the band structure by modifying the on-site energies $\epsilon_j$ with $j = 1, 2$, and $3$, as well as the hopping parameters $t$, $t'$, and $h$, see Fig. \ref{Band_struct}. This modulation of the parameters can be performed through artificial systems straightforwardly, among them photonic and phononic lattices \cite{Zhang2022c,Zhu2023}. Waveguides and elastic resonators in these lattices play the role of artificial atoms, while evanescent coupling mimics bonds among sites \cite{MartinezArgueello2022}. 

In the continuum approximation, we expand the Hamiltonian in Eq. \eqref{HSSSH2} around the symmetry point $K = \frac{\pi}{a}$. With the definition $k = q + K$ and considering linear terms in $q$, we obtain

\begin{equation}
    H(q) = \hbar v_F S_x q + \tilde{M},
\end{equation}

\noindent where the matrix $\tilde{M}$ is defined by

\begin{equation}
     \tilde{M}= \left(\begin{array}{ccc}
        \epsilon_1 & i(t'-t) & h \\
        -i(t'-t) & \epsilon_2 & 0 \\
        h & 0 & \epsilon_3
    \end{array}\right)
\end{equation}

\noindent with Fermi velocity $v_F = a(t + t')/(2\hbar)$. 

If we consider a position-dependent staggered potential $V(x)$ in the bearded SSH lattice, it is necessary to establish the continuity conditions of the wave function in the interface. For instance, the case of a bipartite chain with a step potential and interface located at $x = 0$. To find this matching condition, we integrate the differential equation given by
\begin{widetext}
\begin{eqnarray}
\lim_{\epsilon \rightarrow 0}\int\limits^\epsilon_{-\epsilon} [H(q) + V(x)]\vec{\Psi}(x) dx & = &  E\lim_{\epsilon \rightarrow 0}\int\limits^\epsilon_{-\epsilon}\vec{\Psi}(x) dx\\
    \lim_{\epsilon \rightarrow 0}\int\limits^\epsilon_{-\epsilon}\left(\begin{array}{ccc}
        \epsilon_1+V(x) & -i\hbar v_F\frac{d}{dx}+i(t'-t) & h \\
        -i\hbar v_F\frac{d}{dx}-i(t'-t) & \epsilon_2 +V(x)& 0 \\
        h & 0 & \epsilon_3+V(x)
    \end{array}\right)\left(\begin{array}{c}
    \psi_1(x)\\
    \psi_2(x)\\
    \psi_3(x)\end{array}\right)dx & = & E \lim_{\epsilon \rightarrow 0}\int\limits^\epsilon_{-\epsilon}\left(\begin{array}{c}
    \psi_1(x)\\
    \psi_2(x)\\
    \psi_3(x)\end{array}\right) dx.
\end{eqnarray}
\end{widetext}

\noindent We assume that there are no singularities for the potential and the wave function. In this way, it is only possible to establish the continuity of two components $\psi_1(x)$ and $\psi_2(x)$ because the third component $\psi_3(x)$ does not have continuity necessarily.

\section{Transmission in bearded SSH lattice junctions}

\begin{figure}
    \centering
    \begin{tabular}{c}
    (a) \qquad \qquad \qquad \qquad \qquad \qquad \qquad \qquad \qquad \qquad \qquad \qquad \\ 
      \includegraphics[scale=0.3]{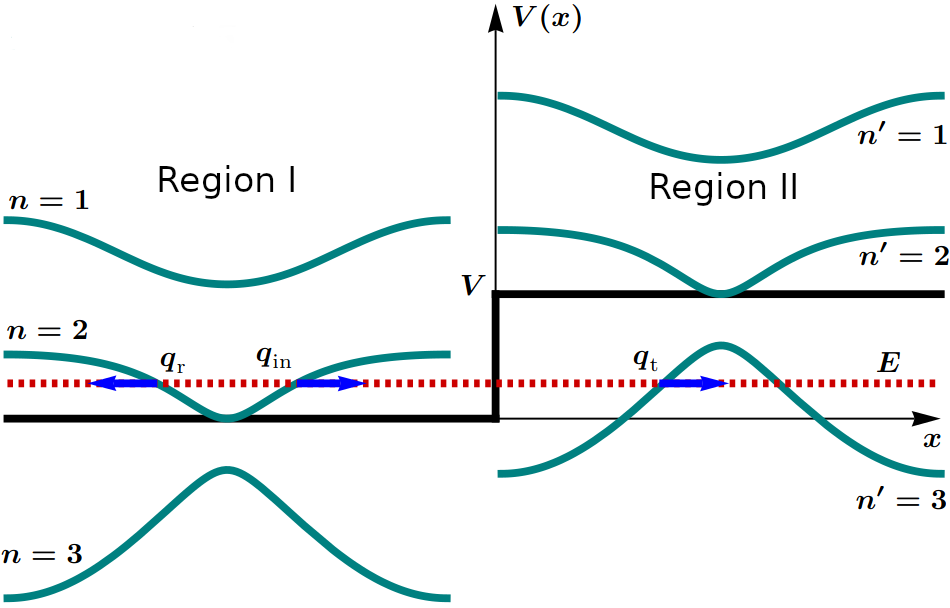}\\
      \\
      (b) \qquad \qquad \qquad \qquad \qquad \qquad \qquad \qquad \qquad \qquad \qquad \qquad \\ 
      \includegraphics[scale=0.3]{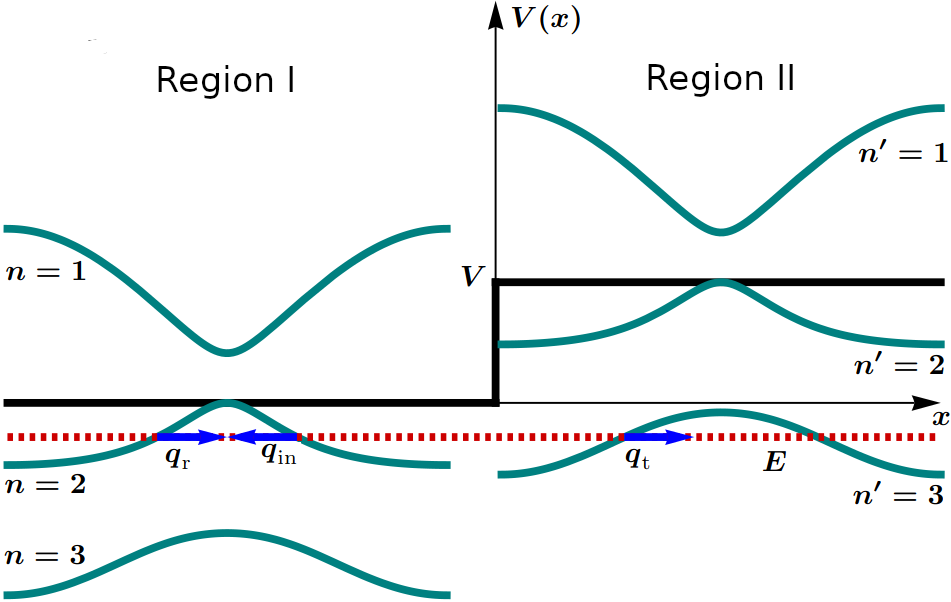}
      \end{tabular}
    \caption{Electronic band structure of a $pn$ junction (green curves) based on bearded SSH lattice. This bipartite lattice emerges by applying different potentials in regions I and II. The blue arrows represent the wave vectors $q_\textrm{in}$, $q_\textrm{r}$, and $q_\textrm{t}$ in the electron scattering. (a) Energy bands with $\epsilon_1 = \epsilon_2 = 0$, $\epsilon_3 = 0.5$, $t = t' = 1$, and $V = 1$. The Fermi level $E$, which is indicated by the dashed red line, shows the interband transmission of electrons from the middle band in region I to the valence band in region II. (b) Same as in (a), but $\epsilon_3 = -0.5$, the flat band is bent for obtaining negative curvature in the center.}
   \label{chain_junc}
\end{figure}
We propose a bipartite bearded SSH lattice that consists of two coupled semi-infinity bearded SSH chains, which are modeled by a step potential. The regions I ($x<0$) and II ($x\geq 0$) have the electrostatic potential 0 and $V$, respectively. The wavefunctions for both regions are written in terms of the eigenvectors of Hamiltonian in Eq. \eqref{HSSSH}

\begin{equation}
    \vec{\Psi}_\textrm{I}(x) = \vec{u}(q_\textrm{in})\textrm{e}^{iq_\textrm{in}x} + {A_\textrm{r}}\vec{u}(q_\textrm{r})\textrm{e}^{iq_\textrm{r}x}, 
\end{equation}
\noindent for the region I and

\begin{equation}
    \vec{\Psi}_\textrm{II}(x) = A_\textrm{t}\vec{u}(q_\textrm{t})\textrm{e}^{iq_\textrm{t}x} 
\end{equation}

\noindent is the transmitted wavefunction in region II. The quantities $A_\textrm{r}$ and $A_\textrm{t}$ are the wave amplitudes, while $q_\textrm{r}$ and $q_\textrm{t}$ are the wave vectors. We evaluate the wave vector $q_\textrm{in}$ as well as $q_\textrm{r}$ and $q_\textrm{t}$ in the eigenvector $\vec{u}(q_{\textrm{in}/\textrm{r}/\textrm{t}})$ according to the schematic representation of the electron scattering in the bipartite bearded SSH lattice in Fig. \ref{chain_junc}. We can select interband transmission, which corresponds to the tunneling of a single electron from the band $n$ in region I to other one $n'$ in region II. Another possibility is the intraband transmission with $n = n'$. We define the notation $n$-$n'$ for the transmission to indicate that the incident electron comes from the band $n$ in region I and crosses to the band $n'$ in region II. Propagation modes occur if electrons with energy $E$ from the band $n$ cross to the band $n'$ when there are allowed states. We note that the group velocity direction depends on the slope sign of the tangent straight to the band. With positive curvature near to $q = 0$, the maximum and minimum are located at $q_{\textrm{in},\textrm{t}}a = \pm \pi$ and $q_{\textrm{in},\textrm{t}}a = 0$, respectively, and group velocity is parallel to the wave vector, as shown in Fig. \ref{chain_junc}(a). The opposite case occurs for negative curvature in the center of the band, as shown in Fig. \ref{chain_junc}(b). Propagation modes emerge if electrons from the band $n$ impinging at the interface have an energy $E_n(q_\textrm{in}a)$ which lies on the overlap between the ranges $\left[\textrm{min}\{E_n(0),E_n(\pi))\}, \max\{E_n(0),E_n(\pi)\}\right]$ and $\left[\min\{E_{n'}(0),E_n'(\pi)\}+V, \max\{E_{n'}(0),E_{n'}(\pi)\}+V\right]$.

Applying the continuity condition at $x = 0$ in the first and second components of $\Psi_\textrm{I}(x)$ and $\Psi_\textrm{II}(x)$, we have the $2 \times 2$ linear equation system

\begin{equation}\label{syst}
    \left(\begin{array}{c}
         u^{(1)}(q_\textrm{in})\\
          u^{(2)}(q_\textrm{in})
    \end{array}\right) + A_\textrm{r}\left(\begin{array}{c}
         u^{(1)}(q_\textrm{r})\\
          u^{(2)}(q_\textrm{r})
    \end{array}\right) = A_\textrm{t}\left(\begin{array}{c}
         u^{(1)}(q_\textrm{t})\\
          u^{(2)}(q_\textrm{t})
    \end{array}\right).
\end{equation}

\noindent Assuming that there is no breaking of the reflection law $q_\textrm{r} = -q_\textrm{in}$, the reflection coefficient is

\begin{eqnarray}\label{Refl}
    R(q_\textrm{in}) & = & |A_\textrm{r}|^2 \nonumber\\
    & = & \left|\frac{u^{(1)}(q_\textrm{in})u^{(2)}(q_\textrm{t})-u^{(2)}(q_\textrm{in})u^{(1)}(q_\textrm{t})}{u^{(1)}(q_\textrm{t})u^{(2)}(-q_\textrm{in})-u^{(2)}(q_\textrm{t})u^{(1)}(-q_\textrm{in})}\right|^2. \nonumber\\
    & & 
\end{eqnarray}

\noindent and the transmission coefficient is given by
\begin{equation}\label{Transm}
    T(q_\textrm{in}) = 1 - R(q_\textrm{in}) 
\end{equation}

\noindent due to conservation of the current density $J_x$. 

\section{Results and discussion}

\begin{figure*}
    \centering
    \begin{tabular}{cccc}
    (a) \qquad \qquad \qquad \qquad \qquad \qquad \qquad \qquad & & (b) \qquad \qquad \qquad \qquad \qquad \qquad \qquad \qquad & (c) \qquad \qquad \qquad \qquad \qquad \qquad \qquad \qquad\\
    \includegraphics[scale=0.22]{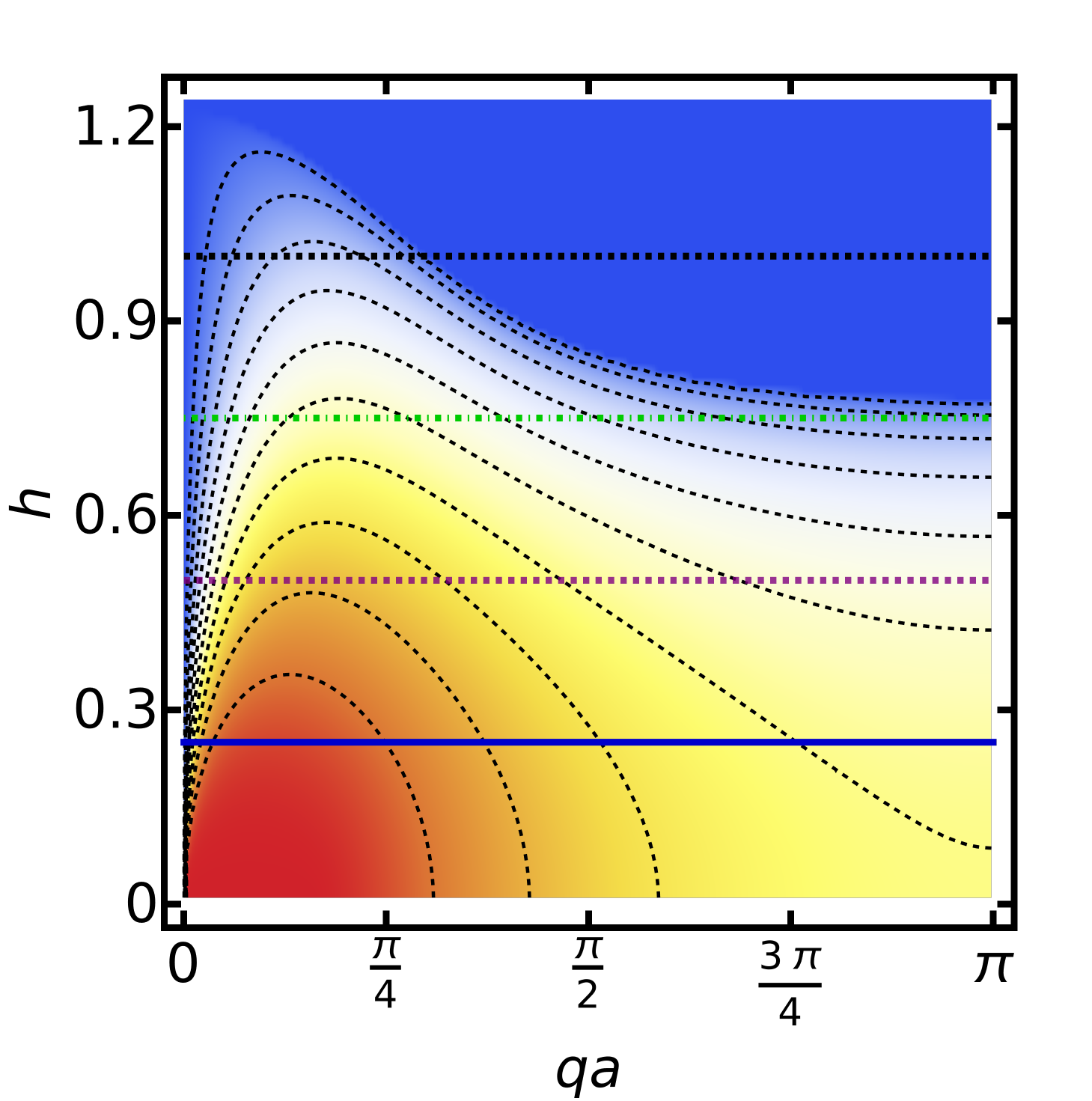} &
    \includegraphics[width=0.8cm,height=5.2cm]{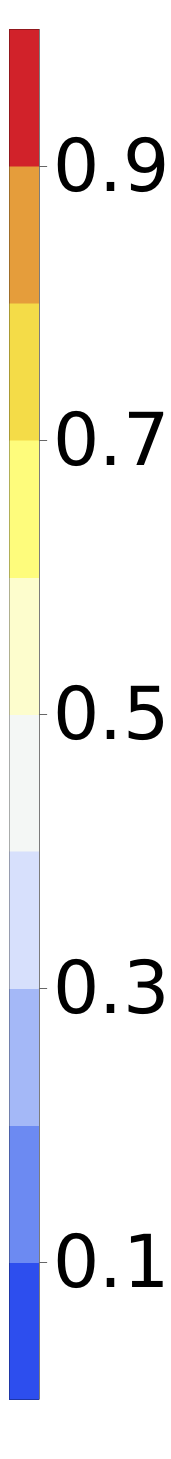} &
    \includegraphics[scale=0.22]{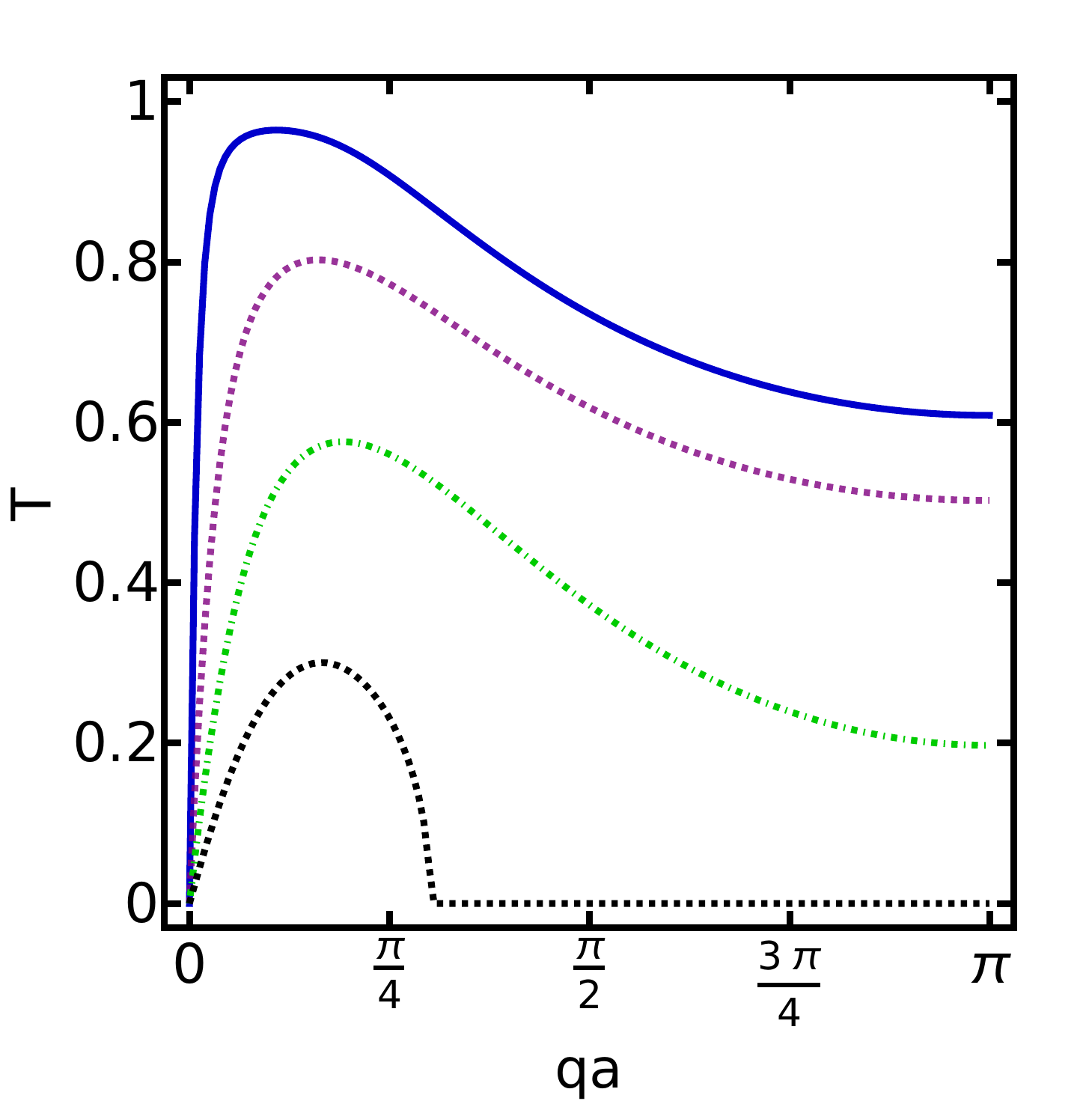} &
    \includegraphics[scale=0.22]{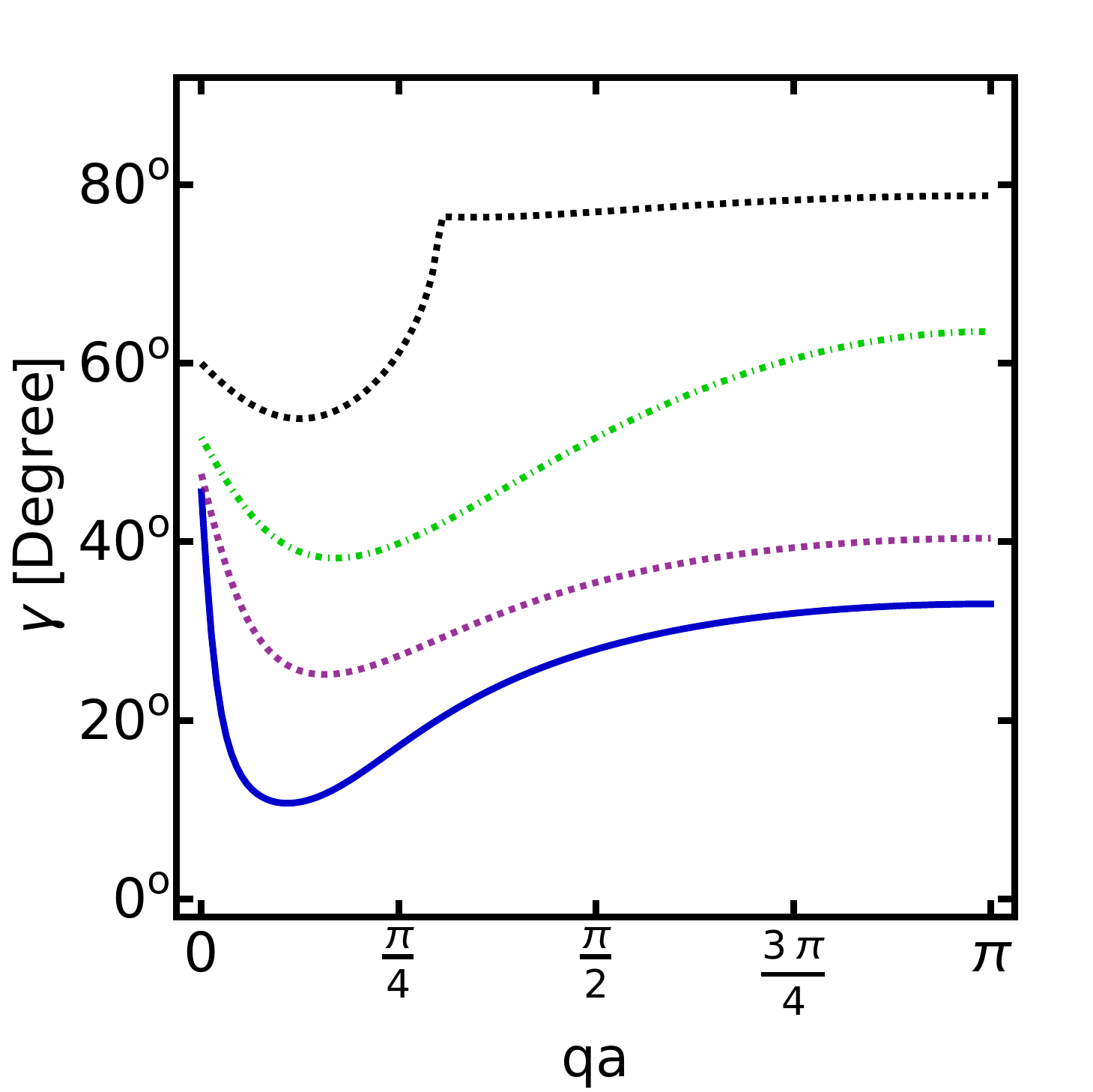}
    \end{tabular}
    \caption{(a) Interband transmission from the middle band $n = 2$ to valence band $n' = 3$ as a function of the hopping parameter $h$ and wave vector $q$ by using the set of values $\epsilon_1 = \epsilon_2 = 0$, $\epsilon_3 = 0.5$, $t = t' = 1$, and potential $V = 1$ (in $t$ units). Reddish region corresponds to zones with high transmission, while bluish zones indicate low transmission when energy is near a gap. (b) Interband transmission curves as a function of $q$ for the values of the hopping parameter $h = 0.25, 0.5,0.75$, and 1, which are depicted as blue, dashed red, dot-dashed green, and black dashed lines, respectively. These curves have their counterpart as horizontal lines in (a). (c) Relative pseudo-spin angle $\gamma$ as a function of wave vector $q$ indicating that to minimum $\gamma$, high transmission is obtained.}
    \label{fig:transmDP}
\end{figure*}
Our main objective is to select the interband transmissions from the bent flat band in region I towards the valence band in region II  by tuning the on-site energy $\epsilon_3$, as shown in Fig. \ref{chain_junc}. We solve numerically Eqs. \eqref{Refl} and \eqref{Transm} taking into account that the transmitted momentum $q_\textrm{t}$ is determined by finding the root of $E_n(q_\textrm{in}) = E_{n'}(q_\textrm{t}) + V$, with $n = 2$ for the middle band in region I and $n' = 3$ for the valence band in region II. The relation between $q_\textrm{in}$ and $q_\textrm{t}$ is equivalent to the Snell law in quasi- one-dimensional lattices. We analyze two situations that correspond to the change of sign in the curvature of the middle band near to $q = 0$, as shown in Fig. \ref{chain_junc}(a) and (b). If $\partial^2 E_n/\partial q^2 > 0$ at $q = 0$,  wave propagation occurs from left to right for positive values of $q_\textrm{in}$. In the opposite case, the wave vector has to change its sign to satisfy the conservation of the current density $J_x$, as occurs in metamaterials. 

\begin{figure}
    \centering
    \includegraphics[scale=0.3]{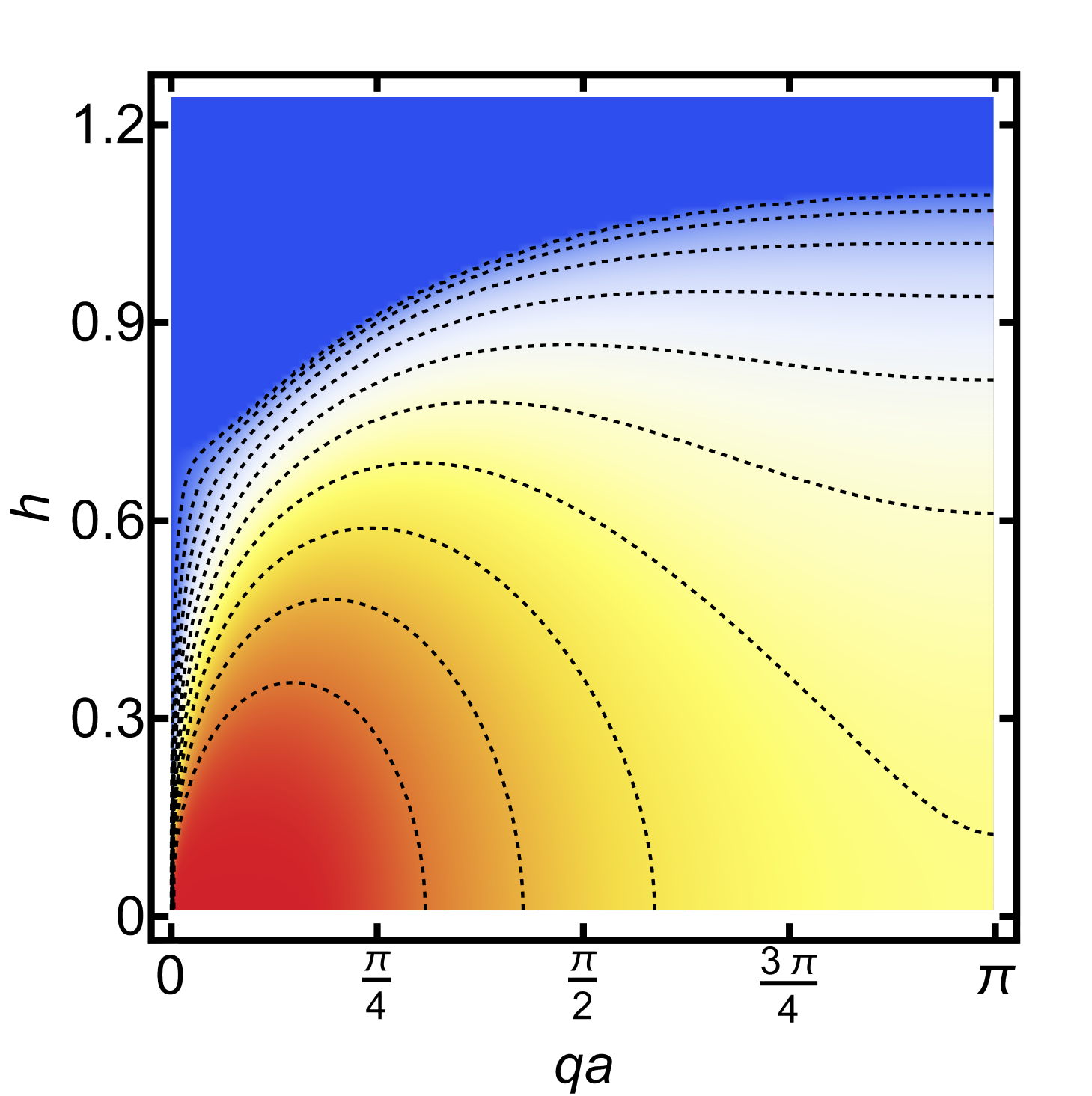} 
    \includegraphics[width=0.9cm,height=7.1cm]{Barra.pdf} 
    \caption{Same as in Fig. \ref{fig:transmDP}, but now $\epsilon_3=-0.5$. Both figures show similar features. In this case, the curvature is negative in the center of the middle band.}
    \label{fig:trams23DPnegcurv}
\end{figure}

\begin{figure*}[t!!]
    \centering
    \begin{tabular}{ccc}
    (a) \qquad \qquad \qquad \qquad \qquad \qquad \qquad \qquad \qquad & & (b) \qquad \qquad \qquad \qquad \qquad \qquad \qquad \qquad \qquad\\
    \includegraphics[scale=0.26]{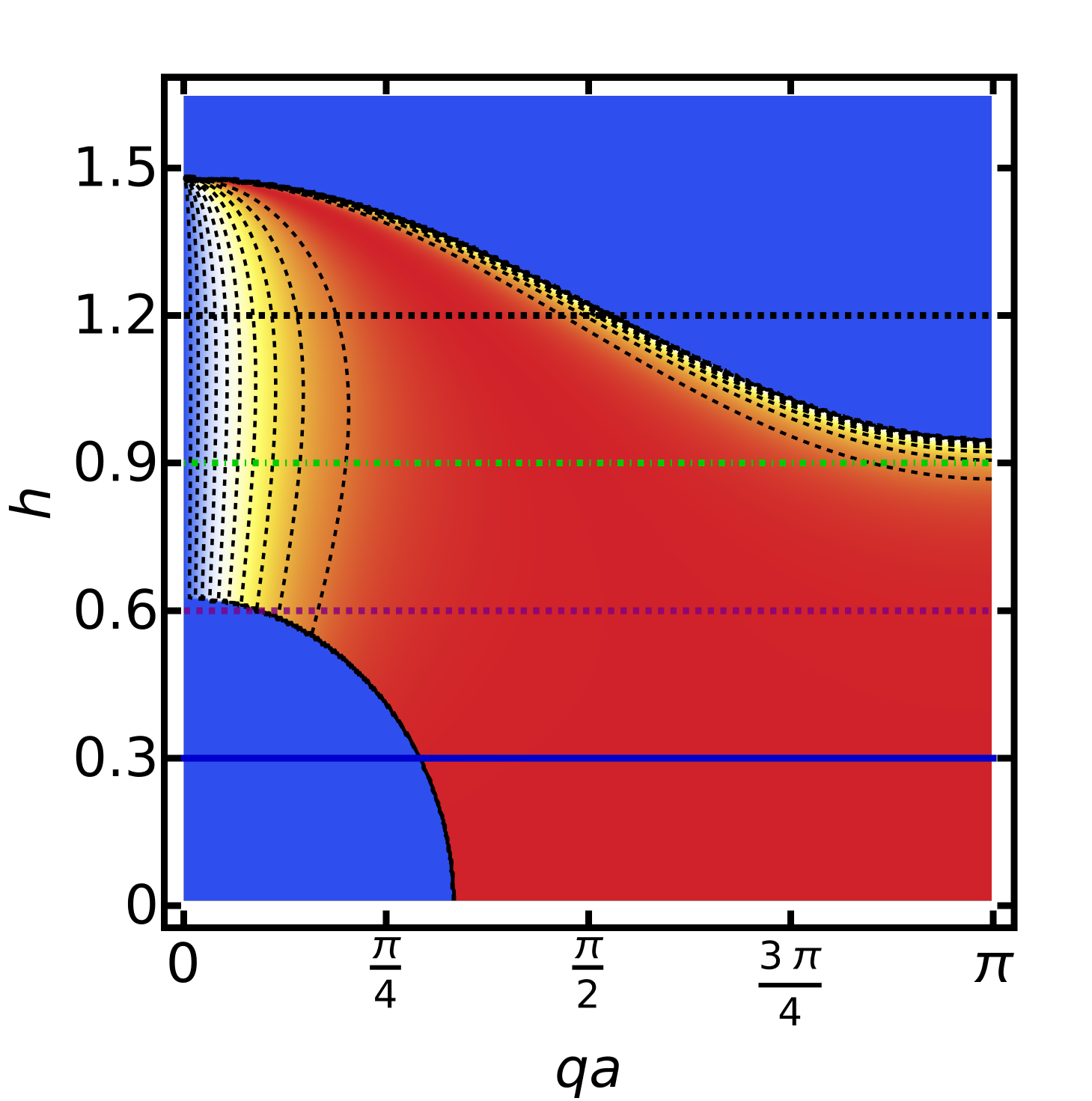} &
    \includegraphics[width=0.8cm,height=6.2cm]{Barra.pdf} &
    \includegraphics[scale=0.26]{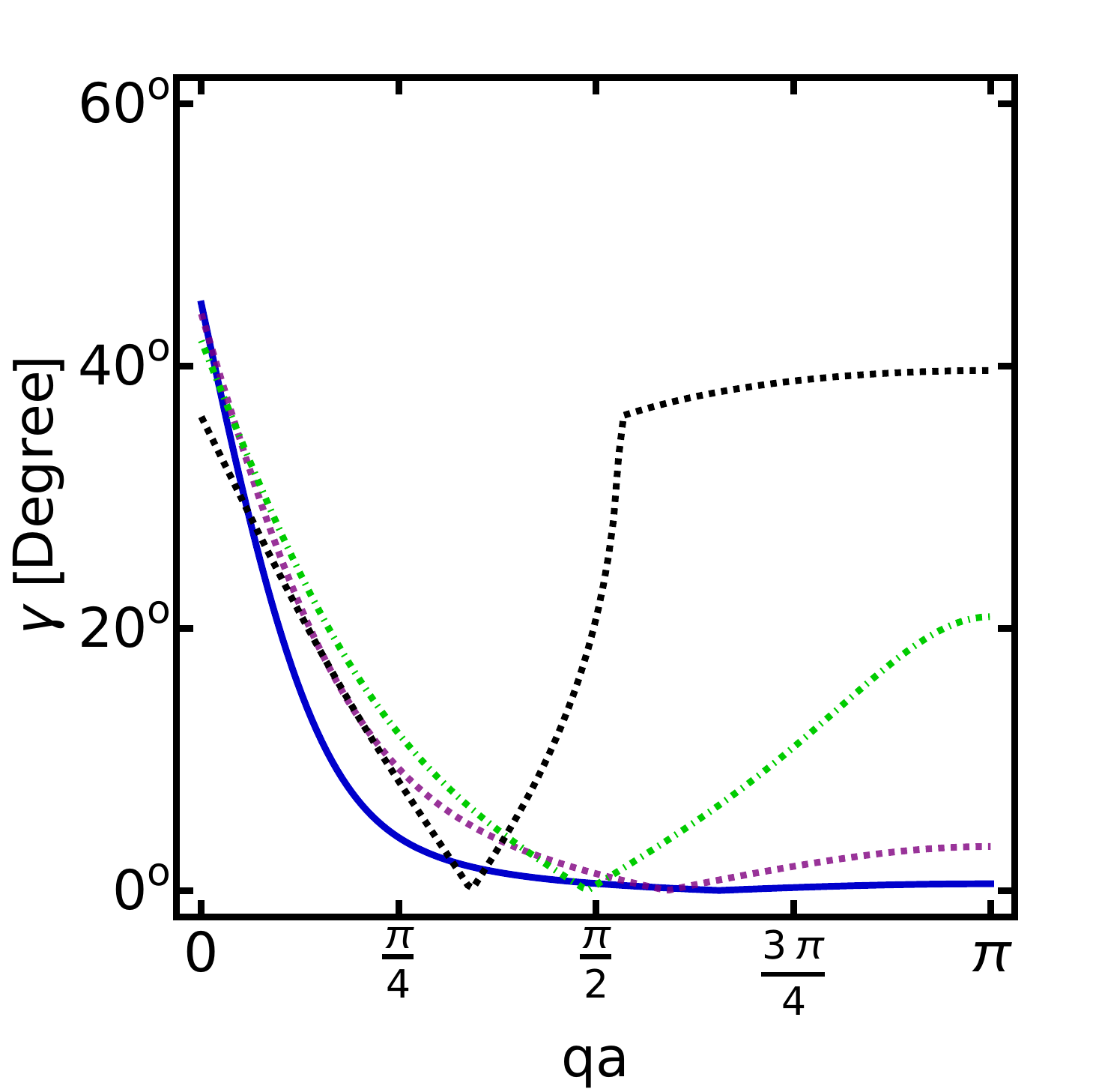}\\
    (c) \qquad \qquad \qquad \qquad \qquad \qquad \qquad \qquad \qquad & & (d) \qquad \qquad \qquad \qquad \qquad \qquad \qquad \qquad \qquad\\
    \includegraphics[scale=0.26]{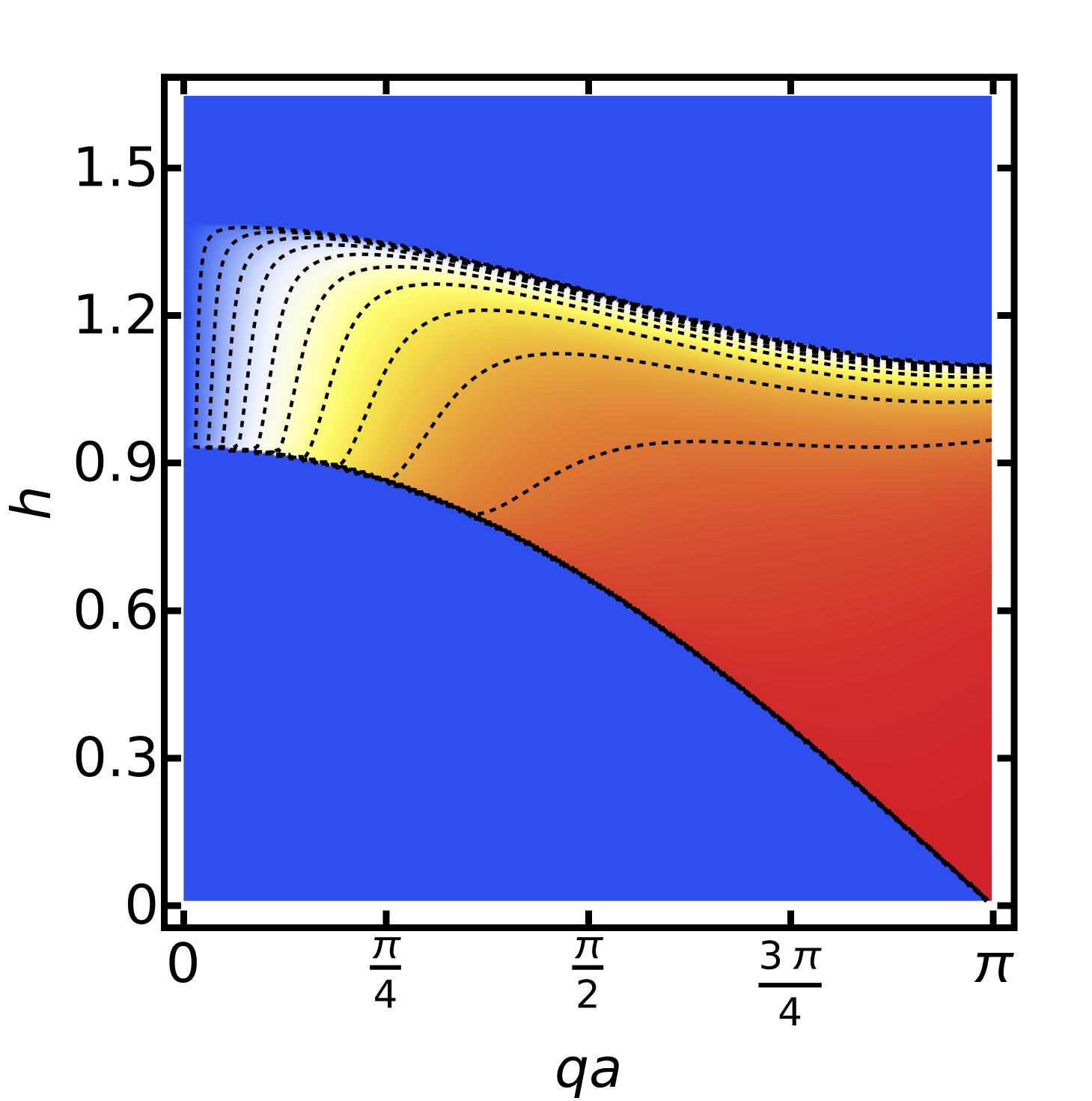} &
    \includegraphics[width=0.8cm,height=6.2cm]{Barra.pdf} &
    \includegraphics[scale=0.26]{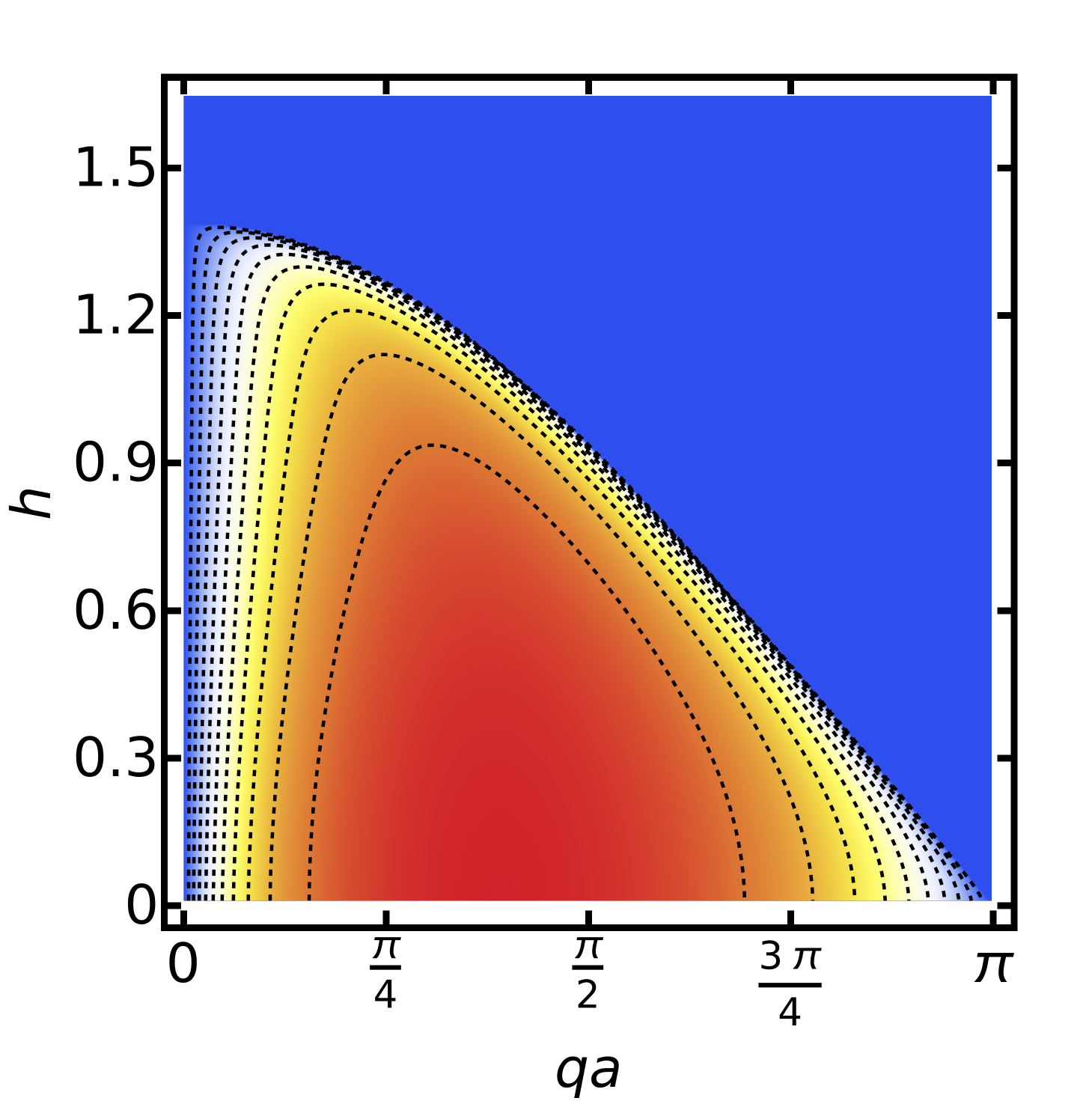}
    \end{tabular}
    \caption{(a) Similar to Fig. \ref{fig:transmDP} but now the transmission is from conduction band $n = 1$ to valence band $n' = 3$ with potential $V = 3$. Klein tunneling is predominant because the pseudo-spin of electrons in bearded SSH lattices is one. (b) Relative pseudo-spin angle as a function of $q$ shows a decay when $h$ decreases. Colors and line types for the curves correspond to the values of $h = 0.3, 0.6, 0.9$, and $1.2$, in (a). (c) Interband transmission 1-3 with hopping parameters $t = 1$ and $t' = 0.5$ favors the emergence of evanescent modes for the trivial phase in the SSH model. (d) Interband transmission with $t = 1$ and $t' = 1.5$ for the topological phase.}
    \label{fig:transm13}
\end{figure*}

We set the on-site energies $\epsilon_1 = \epsilon_2 = 0$ and $\epsilon_3 = 0.5$ to bend the middle band, as seen in Fig. \ref{chain_junc}(a). We consider the interband transmission between this middle band with $n = 2$ to the valence band with $n' = 3$. One can bend the band in the opposite direction by setting the value of $\epsilon_3 = -0.5$ (see Fig. \ref{chain_junc}(b)). To analyze the effect of the hopping parameter $h$ on the transmission, we showed the transmission as a function of the momentum $q_\textrm{in}a \equiv qa$ and the hopping parameter $h$ in Fig. \ref{fig:transmDP}. To identify possible signatures of Klein tunneling, we look for areas with red color in the transmission. High transmission appear if the wave vector is within the range $0 < qa < \pi/4$ and the hopping parameter in $0 < h < 0.3$. Particular transmission curves are shown in Fig. \ref{fig:transmDP}(b) for the set of values of $h = 0.25, 0.5$, $0.75$, and 1,  noting that the transmission decreases when the hopping parameter $h$ increases. The transmission features are explained generally by using the pseudo-spin direction. In graphene, Klein tunneling emerges due to the conservation of the pseudo-spin in the normal direction \cite{Katsnelson2006,Young2009}. Though the system analyzed here is quasi-one-dimensional, the orientation of the pseudo-spin in the chain is represented by the polar angles $\theta$ and $\phi$ in the Bloch sphere. Since the electron scattering involves reduced components of the spinors as indicated in Eq. \eqref{syst}, we can relate these spinors in the Bloch sphere for the incident, reflected, and transmitted states with the set of angles $(\theta_\textrm{in}, \phi_\textrm{in})$, $(\theta_\textrm{r}, \phi_\textrm{r})$, and $(\theta_\textrm{t}, \phi_\textrm{t})$ as a function of $q_\textrm{in}$. Nevertheless, our interest is to get the relative angle between the incident and transmitted state in Eq. \eqref{syst}. The calculation of this relative pseudo-spin angle $\gamma$ between the reduced incident and the transmitted states $\vec{w}_i(q_\textrm{in})=(u^{(1)}(q_\textrm{in}),u^{(2)}(q_\textrm{in}))$ and $\vec{w}_t(q_\textrm{t})=(u^{(1)}(q_\textrm{t}),u^{(2)}(q_\textrm{t}))$ is performed using the expression

\begin{equation}\label{dphi}
    \gamma = \arccos\left[\frac{|\langle w_i| w_t \rangle|}{\sqrt{\langle w_i|w_i\rangle\langle w_t|w_t\rangle}}\right].
\end{equation}

\noindent It is important to note that the electrons in the bearded SSH lattice are equivalent to particles having pseudo-spin one, as seen in the Bloch Hamiltonian in Eq. \eqref{HSSSH2}. In pseudo-spin one systems such as $\alpha$-$\tau_3$ and Lieb lattices, super-Klein tunneling appears. One simple explanation of this transmission effect is that the scattering at the interface imposes the reduction of the pseudo-spin to $1/2$, as noted in Eq. \eqref{syst}. The relative difference in the pseudo-spin angle in Eq. \eqref{dphi} indicates that perfect tunneling emerges when $\gamma = 0$, such as Klein tunneling in graphene \cite{Katsnelson2006}. We show this difference in the pseudo-spin direction as a function of $q$ in Fig. \ref{fig:transmDP}(c), where the maximum value in Fig. \ref{fig:transmDP}(b) corresponds to a minimum in $\gamma$. Perfect transmission occurs in $\gamma = 0$, which indicates that the incident and transmitted pseudo-spin are parallel. 

With negative bending of the flat band, the interband transmission 2-3 has slight modifications concerning the positive case (see Figs. \ref{fig:transmDP}(a) and \ref{fig:trams23DPnegcurv} for comparison). However, the involved scattering process presents several differences. For instance, the group velocity has an opposite direction with the wave vector for both regions. Although this transmission is interband, the scattering is identical to the one intraband transmission, where the electron from the valence band in region I transits to the other valence band in region II. Similar transmission and relative pseudo-spin angles are obtained (not shown).

We analyze the interband transmission 1-3 for electrons from the conduction band to the valence band (see Fig. \ref{fig:transm13}). There is a wide range for the wave vector $q$, where Klein tunneling emerges, as shown in Fig. \ref{fig:transm13}(a). When $qa \rightarrow 0$, the transmission is almost zero because the energy is near the band gap. Increasing $qa$, we can observe that the transmission is perfect because the relative pseudo-spin angle $\gamma$ is zero, indicating the conservation of the pseudo-spin. The hopping parameter $h$ modulates the emergence of Klein tunneling, as shown in Fig. \ref{fig:transm13}(b). Other analyses showed the robustness of the transmission by changing the on-site energy $\epsilon_3$, which is the parameter responsible for the bending at the middle band. We found that the change of the  interband transmission 1-3 from a positive to negative curvature around the center of the band is small and similar to the one shown in Fig. \ref{fig:transm13}(a). Such transmission features can be understood by the fact that the electron in the bearded SSH lattice behaves like a particle of pseudo-spin one, as described in Hamiltonian of Eq. \eqref{HSSSH}, which is the quasi- one-dimensional version of the spin-orbit Hamiltonian of the Lieb lattice. The Klein tunneling observed here is robust because it is the reminiscence of the super-Klein tunneling of pseudo-spin one particles \cite{Bercioux2017,BetancurOcampo2017,ContrerasAstorga2020}.

The interband tunneling 1-3 is almost unaffected by the bending of the middle band when $t = t' = 1$. In contrast, if we set the hopping parameters $t = 1$ and $t' = 0.5$ which corresponds to the trivial phase of the SSH model, the transmission changes drastically as seen in Fig. \ref{fig:transm13}(c). The topological phase $t = 1$ and $t' = 1.5$ presents multiple differences in transmission concerning the previous case (see Fig. \ref{fig:transm13}(d)). Thus, Klein tunneling is strongly affected and sensible by tuning the hopping parameters $t$ and $t'$. The reason is due to that $t$ and $t'$ cause a relative shift between the band $n$ and $n'$ still with the same step potential. Such a relative shift induces the emergence of evanescent modes, as seen in Fig. \ref{fig:transm13}(c), there is a boundary that separates the regions with propagation and evanescent modes. From the case $t = t'= 1$, the relative shift of bands is negligible because the on-site energy $\epsilon_3$ controls the curvature of the middle band.

\section{Conclusions and final remarks}

The bearded edge in SSH chains (see Fig. \ref{Bearded_SSH_lattice}) modulates the band structure and transport properties through the on-site energy $\epsilon_3$ and hopping parameter $h$. The role of these parameters is to bend the flat band in order to explore interband tunneling. Electrons from the bent flat band in region I cross to the valence band in region II, and depending on the chosen parameters, Klein tunneling is obtained. In general, electrons in bearded SSH lattices behave like a pseudo-spin one particle and matching conditions impose the reduction of the spinors to two components in the scattering process. Klein tunneling is due to the conservation of the reduced pseudo-spin from 1 to 1/2, as pointed out in Eq. \eqref{dphi}, Figs. \ref{fig:transmDP}, \ref{fig:trams23DPnegcurv}, and \ref{fig:transm13}. The relative pseudo-spin angle between incident and transmitted states is not exclusive to bearded SSH lattices but also to higher dimension materials and enlarged pseudo-spin. 

We analyzed interband transmissionc 1-3 from the conduction to valence band for the case $t = t'$ by noting that the Klein tunneling is robust against the variation of the on-site energy $\epsilon_3$ and vertical bond parameter $h$ because this effect is the quasi-one-dimensional version of the super-Klein tunneling. However, the interband transmission 1-3 is affected strongly when $t \neq t'$ due to the relative shift of bands. For instance, total internal reflection appears by evanescent modes. An experimental setup with $pn$ junctions of organic molecules may help to test the present results. Other platforms may be used, such as quantum dot arrangements \cite{Kiczynski2022}, photonic crystals \cite{Zhang2022c}, or topolectric circuits \cite{Lee2018,Dong2021,Albooyeh2023}. In classical systems, such as elastic aluminum resonators \cite{MartinezArgueello2022}, Klein tunneling can be implemented through phonons. Emulation of $pn$ junctions can be realized by modifying the resonator size in half of the chain to obtain a bipartite bearded SSH lattice.

\section{Acknowledgments}
We gratefully acknowledge financial support from UNAM-PAPIIT under Project IA106223.

\appendix

\section*{Appendix \\ General tight binding Hamiltonian of a crystal with $n$ atoms in the unit cell}

The Bloch wavefunction of the electron in a crystal which satisfies the Bloch theorem is given by

\begin{equation}\label{BWF}
    \psi_j(\vec{r},\vec{k}) = \sum_{\vec{R}_\ell} \textrm{e}^{i\vec{k}\cdot(\vec{R}_\ell +\vec{\delta}_j)} \phi_j(\vec{r} -\vec{\delta}_j -\vec{R}_\ell),
\end{equation}

\noindent where $\vec{k}$ is the wave vector, the index $j = 1,2,\ldots,n$ and $\vec{R}_\ell$ are lattice vectors, the index $\ell$ labels the unit cell positions. The vectors $\vec{\delta}_j$ correspond to the atomic positions within the unit cell. The function $\phi_j$ is the $j$-th atomic orbital. In the following, we consider only a piece of crystal of $N$ unit cells and we suppose the expression in Eq. \eqref{BWF} remains valid, where $\ell$ goes from 1 to $N$.

In order to get the matrix representation of the TB Hamiltonian, we split the Hamiltonian in two parts

\begin{equation}
    \hat{H} = \hat{H}_\textrm{at} + \hat{V}_\textrm{cr},
\end{equation}

\noindent where $\hat{H}_\textrm{at}$ is the atomic Hamiltonian acting on the site $j$:

\begin{equation}
    \hat{H}_\textrm{at}\phi_j(\vec{r}) = \epsilon_j\phi_j(\vec{r})
\end{equation}

\noindent and $\hat{V}_\textrm{cr}$ describes the interaction of the electron with the whole crystal. The expressions for the matrix elements are

\begin{equation}
    H_{jm} = \int\limits_V \psi^*_j(\vec{r})(\hat{H}_\textrm{at} + \hat{V}_\textrm{cr})\psi_m(\vec{r}) dV.
\end{equation}

\noindent Substituting in this expression, the Bloch wave function of Eq. \eqref{BWF}, we have

\begin{equation}\label{Hjm}
    H_{jm}(\vec{k})=\sum_{\vec{R}'_\ell,\vec{R}_\ell} \textrm{e}^{i\vec{k}\cdot(\vec{R}_\ell -\vec{R}'_\ell + \vec{\delta}_m-\vec{\delta}_j)}\left[\epsilon_j W^{\ell\ell'}_{jm} + T^{\ell\ell'}_{jm}\right],
\end{equation}

\noindent where we define

\begin{equation}
    W^{\ell\ell'}_{jm}(\vec{R}_\ell) = \int\limits_V\phi^*_j(\vec{r}-\vec{\delta}_j-\vec{R}'_\ell)\phi_m(\vec{r}-\vec{\delta}_m-\vec{R}_\ell)dV,
\end{equation}

\noindent which is known as overlap integral because it quantifies the grade of overlapping between the orbital at the site $j$ in the unit cell $\ell$ with one located at the site $m$ within the unit cell $\ell'$. While

\begin{equation}
    T^{\ell\ell'}_{jm} = \int\limits_V\phi^*_j(\vec{r}-\vec{\delta}_j-\vec{R}'_\ell)V_\textrm{cr}(\vec{r})\phi_m(\vec{r}-\vec{\delta}_m-\vec{R}_\ell)dV
\end{equation}

\noindent is the hopping integral and must be interpreted as a probability amplitude of the electron that hops from the atom $j$ in the unit cell $\ell$ to the atomic site $m$ in the unit cell $\ell'$. 

Since the vector $\Delta\vec{R}=\vec{R}_\ell - \vec{R}'_\ell$ is also a lattice vector, there are $N$ repeated terms in Eq. \eqref{Hjm} due to the $N$ unit cells in the crystal. Therefore, the Hamiltonian in Eq. \eqref{Hjm} can be written as

\begin{equation}
    H_{jm}(\vec{k}) = N(\epsilon_jW_{jm}(\vec{k}) + T_{jm}(\vec{k})) = N h_{jm}(\vec{k}),
\end{equation}

\noindent where the overlap matrix elements are

\begin{equation}
    W_{jm}(\vec{k}) = \sum_{\Delta\vec{R}}w_{jm}^{(\Delta\vec{R})}\textrm{e}^{i\vec{k}\cdot(\Delta\vec{R}+\vec{\delta}_{mj})},
\end{equation}

\noindent with 

\begin{equation}
    w_{jm}^{(\Delta\vec{R})} = \int\limits_V \phi^*_j(\vec{r})\phi_m(\vec{r}-\Delta\vec{R}-\vec{\delta}_{mj})dV,
\end{equation}

\noindent and the hopping matrix elements are 

\begin{equation}
    T_{jm}(\vec{k}) = \sum_{\Delta\vec{R}}t_{jm}^{(\Delta\vec{R})}\textrm{e}^{i\vec{k}\cdot(\Delta\vec{R}+\vec{\delta}_{mj})},
\end{equation}

\noindent where

\begin{equation}
    t_{jm}^{(\Delta\vec{R})} = \int\limits_V \phi^*_j(\vec{r})V(\vec{r})\phi_m(\vec{r}-\Delta\vec{R}-\vec{\delta}_{mj})dV.
\end{equation}

An approximation in tight-binding approach is to neglect the overlap among atoms, being $w_{jm}$ approximately the identity matrix elements. 

In the case of three atoms in the unit cell, we have

\begin{equation}
    h(\vec{k}) \approx \left(\begin{array}{ccc}
      \epsilon_1   &  T_{12}(\vec{k}) & T_{13}(\vec{k})\\
      T_{12}^*(\vec{k})  & \epsilon_2 & T_{23}(\vec{k})\\
      T_{13}^*(\vec{k}) & T_{23}^*(\vec{k}) & \epsilon_3
    \end{array}\right)
\end{equation}

\noindent and considering hopping parameters to nearest neighbors

\begin{eqnarray}
    T_{12}(\vec{k})& \approx & t_{12}\textrm{e}^{i\vec{k}\cdot\vec{\vec{\delta}}_{21}} + t'_{12}\textrm{e}^{i\vec{k}\cdot\vec{\vec{\delta}}'_{21}},\\
    T_{13}(\vec{k})& \approx & t_{13}\textrm{e}^{i\vec{k}\cdot\vec{\vec{\delta}}_{13}}\\
    T_{13}(\vec{k})& \approx & t_{23}\textrm{e}^{i\vec{k}\cdot\vec{\vec{\delta}}_{23}}.
\end{eqnarray}

\noindent The first expression for $T_{12}(\vec{k})$ corresponds to the interactions of the atoms at the sites 1 and 2. We show the case when the atom at the site 1 is shared once with the adjacent unit cell, where $t_{12}$ and $t'_{12}$ are the hopping parameters. These can be different by anisotropy in the lattice. For the functions $T_{13}(\vec{k})$ and $T_{23}(\vec{k})$, we considered interactions within the unit cell only with $t_{13}$ and $t_{23}$ the nearest neighbor interactions. In the particular case of the bearded SSH lattice, we set the hopping parameters $t_{12} = t$, $t'_{12} = t'$, $t_{13} = h$, and $t_{23} = 0$ with the relative position vectors $\vec{\delta}_{12} = a_0(-\cos\alpha,\sin\alpha)$, $\vec{\delta}'_{12} = a_0(\cos\alpha,\sin\alpha)$, and $\vec{\delta}_{13} = -a_0(0,1)$, to obtain the Hamiltonian in Eq. \eqref{HSSSH}.

\end{document}